\documentclass[reprint,
  superscriptaddress,  
  amsmath,amssymb,
  aps
]{revtex4-2}

\usepackage{graphicx}
\usepackage{siunitx}

\usepackage{dcolumn}
\usepackage{bm}

\begin{document}

\preprint{APS/123-QED}

\title{Passive freeze-out of the Richtmyer-Meshkov instability}

\author{J. Strucka}
\email{jergus.strucka15@imperial.ac.uk}
\affiliation{%
Plasma Physics Group, Imperial College London, London SW7 2BW, United Kingdom
}%
\author{D. M. Sterbentz}
\affiliation{%
Lawrence Livermore National Laboratory, Livermore, California 94550, USA
}%

\author{B. Lukic}
\affiliation{%
ESRF - The European Synchrotron, CS 40220, 38043 Grenoble Cedex 9, France
}%
\author{K. Mughal}
\affiliation{%
Plasma Physics Group, Imperial College London, London SW7 2BW, United Kingdom
}%
\author{Y. Yao}
\affiliation{%
Plasma Physics Group, Imperial College London, London SW7 2BW, United Kingdom
}%

\author{K. Marrow}
\affiliation{%
Plasma Physics Group, Imperial College London, London SW7 2BW, United Kingdom
}%

\author{W. J. Schill}
\affiliation{%
Lawrence Livermore National Laboratory, Livermore, California 94550, USA
}%

\author{C. F. Jekel}
\affiliation{%
Lawrence Livermore National Laboratory, Livermore, California 94550, USA
}%

\author{D. A. White}
\affiliation{%
Lawrence Livermore National Laboratory, Livermore, California 94550, USA
}%

\author{N. Asmedianov}
\affiliation{%
Physics Department, Technion–Israel Institute of Technology, Haifa 32000, Israel
}%

\author{R. Grikshtas}
\affiliation{%
Physics Department, Technion–Israel Institute of Technology, Haifa 32000, Israel
}%
\author{O. Belozerov}
\affiliation{%
Physics Department, Technion–Israel Institute of Technology, Haifa 32000, Israel
}%
\author{S. Efimov}
\affiliation{%
Physics Department, Technion–Israel Institute of Technology, Haifa 32000, Israel
}%
\author{J. Skidmore}
\affiliation{First Light Fusion Ltd., Oxford Industrial Park, Yarnton OX5 1QU, United Kingdom
}%
\author{A. Rack}
\affiliation{%
ESRF - The European Synchrotron, CS 40220, 38043 Grenoble Cedex 9, France
}%

\author{Ya. E. Krasik}
\affiliation{%
Physics Department, Technion–Israel Institute of Technology, Haifa 32000, Israel
}%

\author{J. L. Belof}
\affiliation{%
Lawrence Livermore National Laboratory, Livermore, California 94550, USA
}%

\author{J. P. Chittenden}
\affiliation{%
Plasma Physics Group, Imperial College London, London SW7 2BW, United Kingdom
}%

\author{S. N. Bland}
\affiliation{%
Plasma Physics Group, Imperial College London, London SW7 2BW, United Kingdom
}%

\date{\today}

\begin{abstract}
The Richtmyer-Meshkov instability (RMI) poses a major challenge in inertial confinement fusion (ICF) due to its role in mixing and performance degradation. We report the first experimental observation of passive freeze-out of RMI in a low-pressure surrogate regime; an instability stagnation effect induced without modifying the driving pressure pulse or the target surface geometry. Using additively manufactured sub-surface voids in a sinusoidal target, we convert a single shock into a sequence of weaker shocks that suppress instability growth upstream of the surface by over $70\%$. High-speed X-ray imaging and hydrodynamic simulations suggest that this suppression arises primarily from temporal shaping, with lesser contributions from spatial curvature and shock weakening. Our results demonstrate a driver-independent pathway for controlling shock-driven hydrodynamic instabilities relevant to ICF and other high energy density systems.
\end{abstract}

\maketitle


\begin{figure*}[!hbt]
\includegraphics[width=\linewidth]{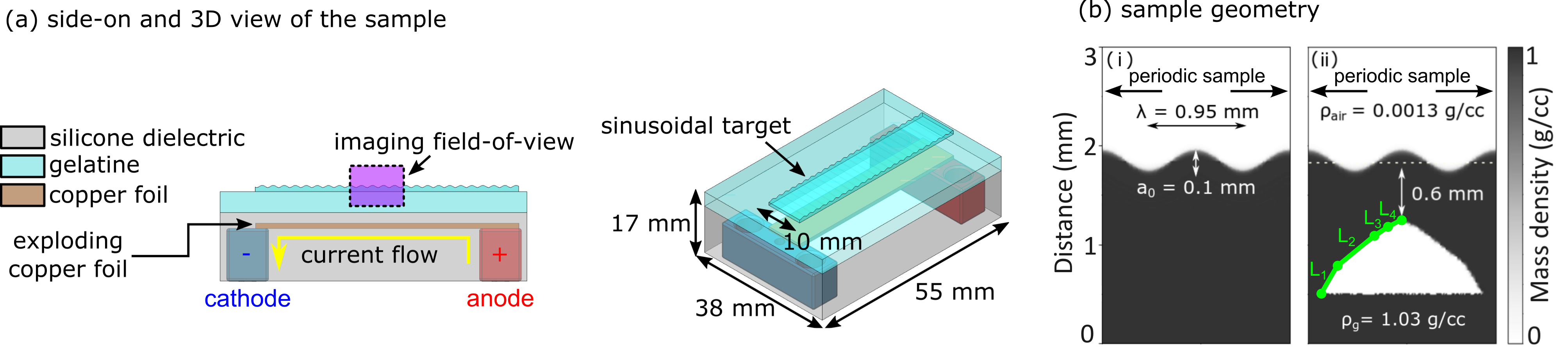}
\caption{\label{scheme}(a) Side-on and isometric view of the experimental setup showing the exploding copper foil, direction of the current flow, and sample geometry. (b) Initial conditions of the baseline (i) and instability mitigation (ii) sample geometry. Functions $L_i$ correspond to linear functions that approximate the optimal void geometry that is detailed in the Methods section.}
\end{figure*}

The Richtmyer-Meshkov instability (RMI) arises when an interface between fluids of differing densities undergoes impulsive acceleration, typically from shock waves \cite{Zhou2017, Zhou2017c}. This triggers dynamics such as interface distortion, jetting, and uncontrollable fluid mixing regardless of the initial density configuration. Controlling RMI is essential for uniformity in compression experiments and inertial confinement fusion (ICF), as recently demonstrated by the success at the National Ignition Facility \cite{Zylstra2022}. In ICF, the instability drives mixing of high-$Z$ materials from outer layer of the fuel capsule into the fusion fuel. Such impurities radiate energy efficiently, lowering the core fuel temperature and ultimately halting the thermonuclear reaction \cite{Keenan2020, Clark2024}.
This letter reports experimental observation of passive freeze-out, a phenomenon based on a prediction first made by Mikhaelian 40 years ago \cite{PhysRevA.31.410}. Namely, we show that the temporal structure of a pressure pulse generated by the shock-sample interaction can self-suppress RMI. This effect opens a new class of alternative strategies aimed at mitigating instabilities, alongside graded density transitions \cite{Sano2020}, reverberation waves across multiple interfaces \cite{Liang2021}, tailored shock wave pairs \cite{Schill2023}, magnetic stabilization in plasmas \cite{Samtaney2003,Sano2013}, and radiative stabilization \cite{Huntington2018}.

While nearly ideal initial conditions would also prevent instability growth, achieving such symmetry is impractical in most scenarios \cite{Haan2011}. For instance, in multi-shell direct-drive ICF, maintaining near-spherical capsule symmetry requires adhesive-filled joints \cite{Randolph2016,Olson2020,Vazirani2024}. In indirect-drive ICF, fill-tubes and foam structures are used to inject and soak deuterium-tritium fuel into the capsule. These necessary features act as hydrodynamic defects which seed instabilities and demand advanced mitigation strategies to limit mixing. Reflecting this ongoing challenge, recent studies have closely examined various aspects of fill-tube dynamics \cite{Kuczek2023,Haines2019,Baker2020,Weber2020} and isolated defects \cite{PhysRevLett.125.055001, 10.1063/1.4818280}.
In a surrogate, low-pressure regime, we demonstrate passive stabilization of a shock-compressed sinusoidal interface through introduction of numerically optimized voids. These structures shape the shockwave without placing constraints on the driving technology, suggesting a route toward mitigating instabilities seeded by fill tubes, joins, and other unavoidable perturbations.

\begin{figure}[!ht]
\includegraphics[width=0.84\linewidth]{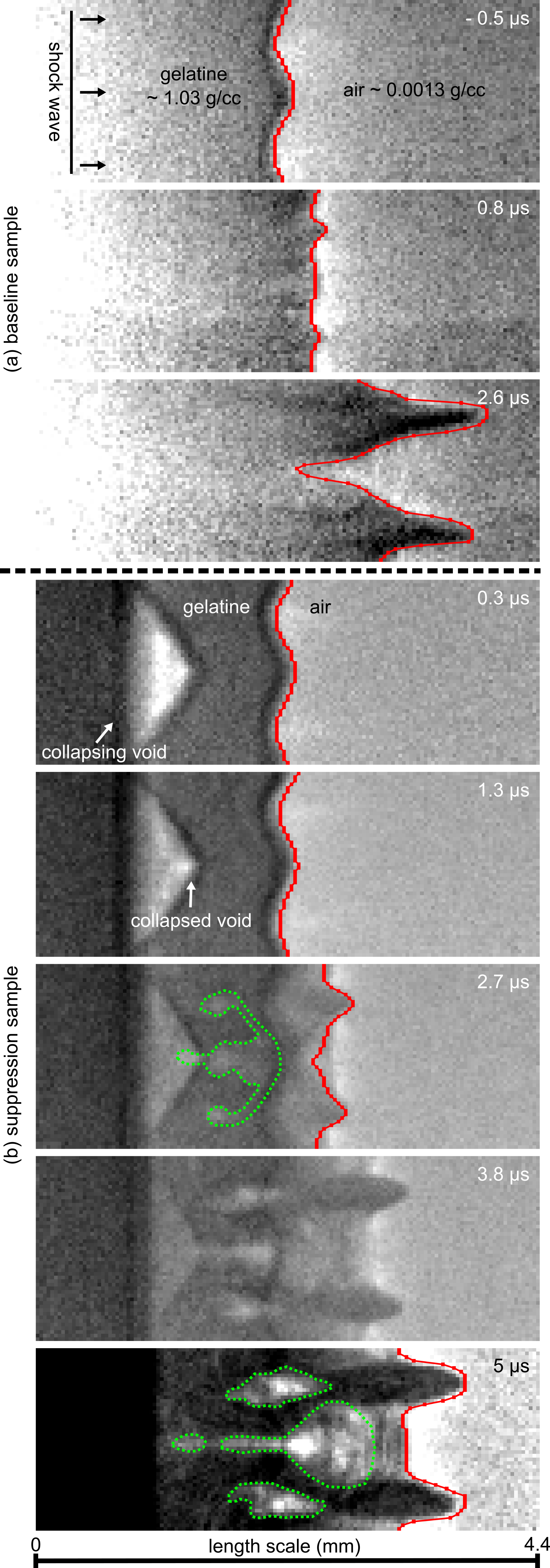}
\caption{\label{fig_exp}(a) X-ray radiographs of our baseline experiment showing shock compression of a sinusoidal surface. (b) Sequence of radiographs showing compression of the instability suppression sample. Green dashed lines highlight void structures while red lines show tracked interfaces.}
\end{figure}

The experiment is illustrated in Fig. \ref{scheme}(a). Two sample geometries were compressed by a shock wave ($P_\text{max} \approx 500~\SI{}{\mega\pascal}$) generated by the electrical explosion of a copper foil: a sinusoidal interface (wavelength $\lambda = 0.95~\SI{}{\milli\meter}$, amplitude $a_0 = 0.1~\SI{}{\milli\meter}$), and a sinusoidal interface with void structures located under the surface for instability suppression. The experiments were performed in air with a density contrast at the perturbed interface given by the Atwood number $A = (\rho_\text{air}-\rho_\text{g}) / (\rho_\text{air}+\rho_\text{g}) \approx -1$, where $\rho_\text{air}$ and $\rho_\text{g}$ represent the densities of air and gelatine, respectively.
Two-dimensional hydrodynamic simulations informed the suppression void design using the trust region Bayesian optimization method \cite{turbo}. The voids were parameterized by four piecewise cubic Hermite polynomials constrained to be within a range of $[0.06, 0.134]~\SI{}{\centi\meter}$ from the interface \cite{Sterbentz2022}. The optimization varied these polynomials to minimize the objective function:
\begin{equation}
\phi = \text{Max}\left(\frac{1}{N_r}\sum^{N_r}_{i=1}\left|h_{\text{p,i}} - h_{\text{b,i}}\right|\right).
\label{eq:opt}
\end{equation}
Here, $h_{\text{p,i}}$ denotes the instability amplitude (``jets"), $h_{\text{b,i}}$ the trough amplitude (``bubbles"), and $N_r$ the number of perturbation periods. Eq. \ref{eq:opt} represents the maximum mixing zone width obtained within the simulation for a specific geometry. To simplify fabrication, the optimal geometry was approximated using four linear segments ($L_1$, $L_2$, $L_3$, and $L_4$). Additional simulations confirmed that this approximation preserved the suppression dynamics. The approximated geometry is shown in Fig.~\ref{scheme}(b, ii). Fig.~\ref{fig_exp} shows the compression and instability dynamics observed using a sequence of time-resolved X-ray radiographs ($\bar{E} = 30~\SI{}{\kilo\electronvolt}$, pulse length $\sim60~\SI{}{\pico\second}$) at the European Synchrotron. More details about the setup and simulations are provided in the appendices. 

 Fig.~\ref{fig_exp}(a) shows the baseline RMI. The driving shockwave propagated through the gelatine sample at $v_\text{shock} = 2.2 \pm 0.2~\SI{}{\kilo\meter\per\second}$ (Mach $1.5$). By $t = 2.6~\SI{}{\micro\second}$, the sinusoidal interface was compressed, inverted, and generated strong jetting, consistent with dynamics expected at highly negative Atwood numbers and previously observed in laser experiments \cite{10.1063/1.4764287}. Between $0.5$ and $5~\SI{}{\micro\second}$, the interface of the baseline experiment advanced at $v^\text{mean}_\text{base} = 316 \pm 10~\SI{}{\meter\per\second}$, while the jet tip propagated at $v^\text{jet}_\text{base} = 651 \pm 25~\SI{}{\meter\per\second}$. In comparison, the suppression geometry in Fig.~\ref{fig_exp}(b) exhibits significantly reduced instability and jetting. Three intensity regions are observed: the thick gelatine substrate required for fabrication of the suppression sample ($\sim50~\SI{}{\milli\meter}$ into the page, dark, not present in baseline), the suppression sample ($\sim10~\SI{}{\milli\meter}$, gray), and the overlying air (bright). Between $t=0.3~\SI{}{\micro\second}$ and $1.3~\SI{}{\micro\second}$, increased attenuation within the engineered void indicates its crushing, followed by the emergence of an upward-propagating, two-lobed bubble structure that dominates the dynamics for $t > 2~\SI{}{\micro\second}$. High-resolution radiographs and simulations show that this structure originates from jetting at the void corners. After inversion, upward-directed jets form from the initial troughs of the sinusoidal surface. The interface, defined by X-ray transmission $T < 0.95$, advances at $v^\text{mean}_\text{supp} = 320 \pm 13~\SI{}{\meter\per\second}$, with jet tips propagating at a much reduced speed $v^\text{jet}_\text{supp} = 407 \pm 19~\SI{}{\meter\per\second}$.

An imprint of the initial void appears in all radiographs of the suppression dynamics. This originates from $\sim100~\SI{}{\micro\second}$ X-ray pre-illumination of the LYSO:Ce scintillator, which induces a slowly decaying, spatially non-uniform increased conversion efficiency reflecting the initial transmission rather than a true double exposure. For $t = 2.7~\SI{}{\micro\second}$ and $5~\SI{}{\micro\second}$, true features are highlighted with green dashed lines to distinguish them from the gradually fading imprint.

\begin{figure}
\includegraphics[width=0.96\linewidth]{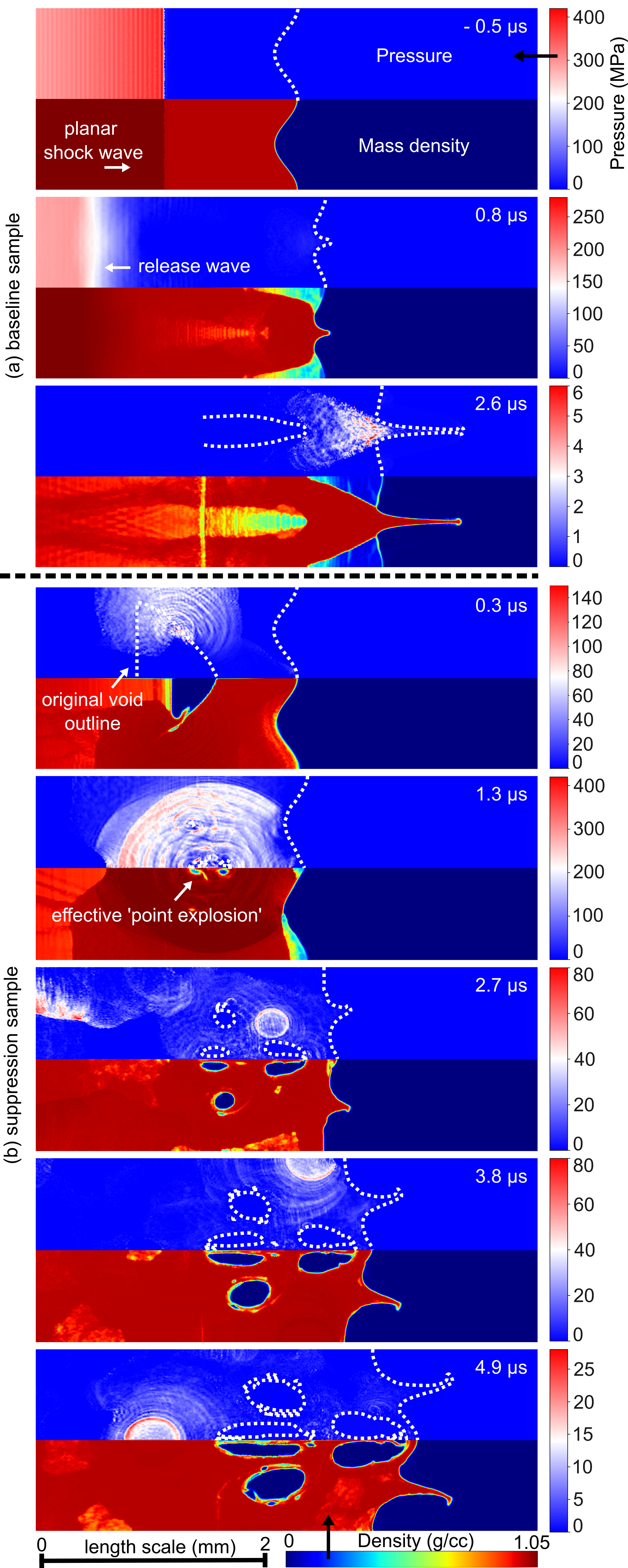}
\caption{\label{fig_3}(a) Simulated pressure and density fields for the baseline scenario. (b) Corresponding fields from the instability suppression case, highlighting key dynamics. In each plot, the top half shows pressure, and the bottom half shows density. Pressure plots include white dashed contours indicating material interfaces or strong density gradients. Note: To highlight features, the pressure color bars vary between plots.}
\end{figure}

\begin{figure*}
\includegraphics[width=0.8\linewidth]{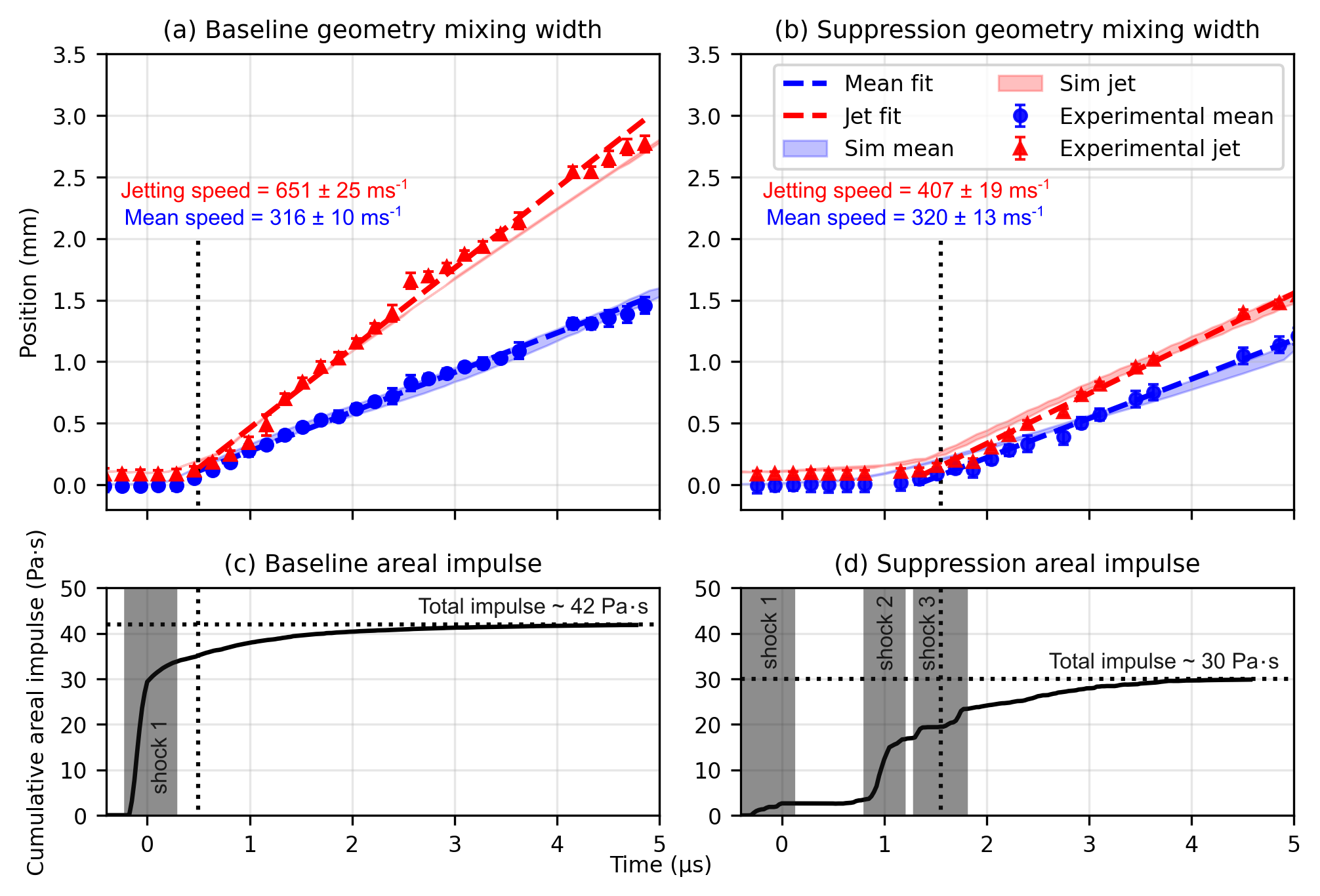}
\caption{\label{fig_4}(a) Jet tip position (red) and mean interface position (blue) vs. time for the baseline geometry.  (b) Same as (a), for the suppression geometry (c) Integrated areal impulse delivered to the gelatine–air interface in the baseline case. (d) Same as (c), for the suppression case. The dashed vertical line indicates the time of interface inversion in all subplots.}
\end{figure*}

There are three mechanisms that can contribute to the observed instability suppression: (1) attenuation of the driving shock wave by the void, (2) temporal shaping of the pressure pulse via conversion of a single strong shock into multiple weaker shocks, and (3) spatial shaping of the shock front, which modifies baroclinic vorticity deposition at the interface through $\nabla\rho \times \nabla p$. Fig.~\ref{fig_3} presents numerical simulations that reproduce the overall dynamics observed in the X-ray radiographs, including void collapse and mixing-zone evolution. Residual differences, such as the apparent jet thickness, are conjectured to arise primarily from line-of-sight integration through the finite sample thickness, alignment, and imperfect surface finish. In the following, we assess the relative contributions of each mechanism using numerical simulations and growth-rate measurements from X-ray imaging (Fig.~\ref{fig_4}).

\textit{Attenuation:} For sufficiently small perturbation amplitudes ($a(t_0)/\lambda \ll 1$), the contribution of shock attenuation to the RMI suppression can be estimated using the linear growth rate \cite{Zhou2017}:

\begin{equation}
\label{RMI}
\frac{da(t)}{dt} = \frac{2\pi}{\lambda} a(t_0) A \Delta u,
\end{equation}
where $a(t)$ and $\lambda$ are the perturbation amplitude and wavelength, $A$ is the Atwood number, and $\Delta u$ is the free surface velocity after shock compression. From Fig.~\ref{fig_exp}, we measure mean position of the interface and its velocity $v^\text{mean}_i$ which corresponds to $\Delta u$ in Eq.~\ref{RMI}. The perturbation growth $da(t) / dt$ corresponds to the jetting speed in the free surface frame given by $v^\text{jet}_i - v^\text{mean}_i$. In linear regime, dividing Eq.~\ref{RMI} by $\Delta u$ yields the ratio $\beta_i = (v^\text{jet}_i - v^\text{mean}_i) / v^\text{mean}_i = 2\pi a(t_0) A /\lambda$ which depends only on the initial geometry. Thus, any reduction in jetting (lower $v^\text{jet}_i - v^\text{mean}_i$) due to shock attenuation (lower $v^\text{mean}_i$) should yield the same $\beta$ parameter. Measurements in Fig.~\ref{fig_4} indicate that $\beta_\text{base} = -1.06 \pm 0.1$ for the baseline geometry and $\beta_\text{supp} = -0.27 \pm 0.08$ for the suppression geometry. While the initial conditions are weakly nonlinear, the large difference in $\beta$ parameters suggests that pure shock attenuation cannot explain the observed reduction in jetting. Nonetheless, numerical simulations show that the areal impulse imparted to the interface decreases from $42~\SI{}{\pascal\cdot\second}$ to $30~\SI{}{\pascal\cdot\second}$ when moving from the baseline to the suppression geometry. This indicates that some, but not all, reduction is due to partial decoupling of the shock from the interface due to the void structure.

\textit{Temporal shaping:} Over 40 years ago, Mikhaelian suggested that shaping the pressure pulse in time, for example by delivering two equally strong planar shocks at two distinct times when the interface amplitude is $a(t_1) = a_0$ and $a(t_2) = -a_0$, can fully suppress linear RMI in negative Atwood number configurations \cite{PhysRevA.31.410}.

Fig.~\ref{fig_3} shows the simulated pressure and density distribution in the baseline and suppression geometries at four key times, while Figs.~\ref{fig_4}(c,d) show the cumulative impulse history experienced by the gelatine-air interface extracted from these simulations. In the suppression case in Fig.~$\ref{fig_3}$, we observe that crushing of the void structure converts the initial pressure profile from a single shock into a sequence of weaker shocks arriving at the interface at various stages of RMI. Simulations show three dominant shocks impacting the interface, the first at $t_1 \approx 0.25~\SI{}{\micro\second}$ with $a\left(t=t_1\right) = 100~\SI{}{\micro\meter}$, $P_1 \approx 35~\SI{}{\mega\pascal}$, and impulse $I_1 = 1.6~\SI{}{\pascal\cdot\second}$. The second follows at $t_2 \approx 1.25~\SI{}{\micro\second}$ with $a\left(t=t_2\right) = 100~\SI{}{\micro\meter}$, $P_2 \approx 255~\SI{}{\mega\pascal}$, and $I_2 = 19.4~\SI{}{\pascal\cdot\second}$; the third at $t_3 \approx 1.75~\SI{}{\micro\second}$ with $a\left(t=t_3\right) = 0$, $P_3 \approx 40~\SI{}{\mega\pascal}$, and $I_3 = 4.88~\SI{}{\pascal\cdot\second}$. Figs.~\ref{fig_4}(c,d) show the corresponding time-resolved areal impulse distributions. The inversion time of the interface, i.e. $a \approx 0$, is delineated by vertical dotted lines.

The effect of this multiple shock structure on the dynamics can be estimated using linear theory in Eq.~\ref{RMI} by treating the total jetting as a linear sum of individual contributions from the three main shocks:

\begin{equation}
\left. \frac{da(t)}{dt} \right|_\text{shocks} = \frac{2\pi}{\lambda} \sum_n a(t_n) A \Delta u_n,
\end{equation}
where $a(t_n)$ is the perturbation amplitude at the time of arrival of the $n$-th shock, and $\Delta u_n = (I_n / I_\text{tot})\Delta u$ is the velocity imparted by each shock, scaled according to its contribution $I_n$ to the total impulse $I_\text{tot}$. Using this model, the jetting speed is predicted to drop from $v^{\text{jet}}_{\text{base}} = 525 \pm 70~\SI{}{\meter\per\second}$ in the baseline scenario to $v^{\text{jet}}_\text{supp} = 465 \pm 50~\SI{}{\meter\per\second}$ in the suppression case, with error bars calculated by combining errors in initial geometry ($\sigma_a = 32~\SI{}{\micro\meter}$, $\sigma_\lambda = 64 ~\SI{}{\micro\meter}$) and mean interface speed $\Delta u$. 

Although simplified, this analysis demonstrates that temporal shaping of shocks interacting with the interface plays a significant role in instability suppression. In practice, the ideal freeze-out may be inaccessible here due to the asymmetric jetting caused by weakly nonlinear initial conditions, and divergent shock waves produced by effective point explosions. As a result, the baroclinic torque and total momentum transfer is unevenly distributed between the ejected jets and the rest of the interface. 

\textit{Spatial shaping:} The development of a non-planar, curved shock front during the shock–sample interaction can influence instability growth by altering the baroclinic vorticity source term, $\nabla \rho \times \nabla p$. This spatial shaping arises from shock propagation around the void, overlapping shock fronts that create doubly shocked regions, and localized void implosions acting as transient point sources of pressure (see Fig.~\ref{fig_3}b). Disentangling spatial effects from temporal shaping is nontrivial. Nevertheless, the simulations show that the dominant pressure pulse, generated by the void crushing and subsequent point explosion at $t \approx 1.3~\SI{}{\micro\second}$, modifies the vorticity deposited across the perturbation. Specifically, it increases baroclinic torque near the simulation boundaries, where the shock impinges at angles nearly perpendicular to the density gradient, and enhances jetting. Conversely, near the center, where the shock curvature aligns with the interface curvature, this shaping leads to weaker vorticity deposition, producing localized stabilization. The net result of these dynamics is curving of the jets away from the central region. Simulations reveal that the vorticity deposition in the central region between the jets in Fig.~\ref{fig_3} is similar across both cases, suggesting that the net effect of spatial shaping in this configuration is not significant.

\begin{figure}[!ht]
\includegraphics[width=0.78\linewidth]{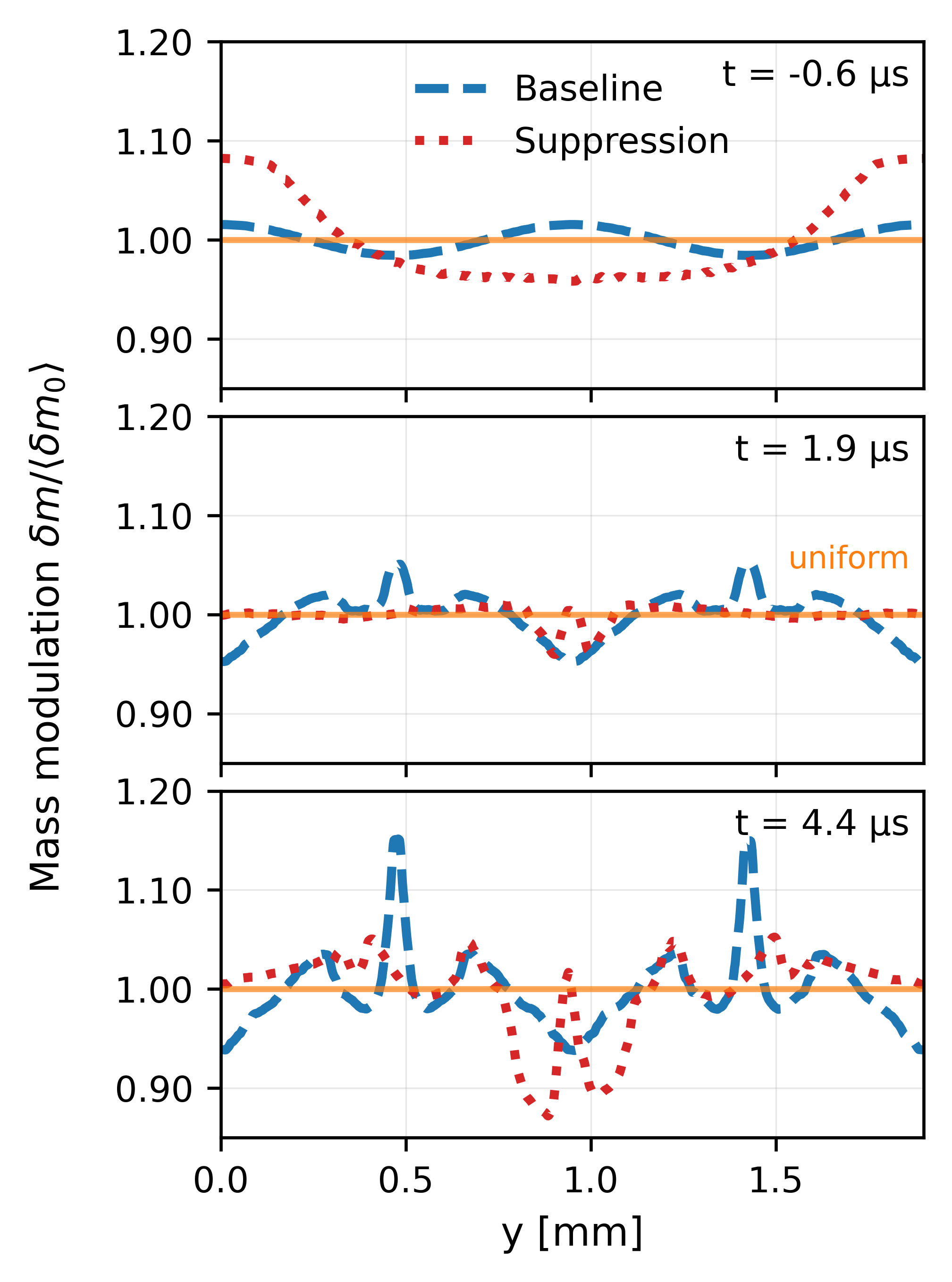}
\caption{\label{mass_modulation} Mass modulation $\delta m / \left \langle \delta m_0 \right \rangle$ at three characteristic times, before shock-compression, shortly after, and at late time. Axis $y$ is perpendicular to direction of compression.}
\end{figure}

It should be noted that, while our optimization focused on the evolution of the mixing zone width due to its direct observability in our experiments, the mass modulation, $\delta m \left(y, t \right)$, defined as the integrated mass density $\rho$ along the axis of compression, will be equally important in more integrated scenarios. In applications such as ICF, the RMI transitions to Rayleigh-Taylor instability (RTI) as the interface decelerates during compression, where the mass modulation plays a critical role in determining the mixing rate and feedthrough \cite{PhysRevLett.87.265001, 10.1063/1.1459459}, and may compete with early-time stabilization. In Figure \ref{mass_modulation}, we evaluate the normalized mass modulation, $\delta m \left(y, t \right) / \left\langle\delta m_0\right\rangle$, from the numerical simulations. Here, $\left\langle\delta m_0\right\rangle$ is the mean areal density at the beginning of the simulation. This modulation is initially higher in the suppression geometry; however, after shock compression, the collapsing voids result in the mass modulation becoming more uniform in the suppression geometry and remains approximately comparable at late time, while simultaneously minimizing long-scale jetting and mix.

In summary, these measurements and simulations demonstrate that growth of the RMI from surface perturbations can be substantially reduced using engineered voids without modifying the surface geometry. In our experiments, we observed suppression of growth in the mixing zone width by over $70\%$ that could be largely explained by temporal shaping of the pressure pulse. Mechanistically, the engineered void acts as a transient energy reservoir, converting a single incident shock into a sequence of pressure pulses—enabling driver-agnostic mitigation of the instability. This approach, alongside novel fabrication methods~\cite{Christ2025,10.1063/5.0107542,10.1063/5.0273572}, suggests a pathway towards controlling instability growth seeded by hydrodynamic features such as fill tubes or adhesive joins, which are present in inertial confinement fusion and high energy density platforms \cite{Haan2011,Randolph2016,Olson2020,Vazirani2024,Kuczek2023,Haines2019,Baker2020,Weber2020}. While the use of shock-cavity interaction to modify pressure distribution appears scalable across multiple pressure regimes~\cite{10.1063/5.0272820,Escauriza2020, Ranjan2011}, the void geometry required for each application will vary and require optimization. The dominant role of temporal shaping in suppression observed within our experiments suggests that this method may be only effective for heavy-to-light interface configurations (negative Atwood number), such as internal walls of ICF capsules. Mitigation under light-to-heavy configurations (positive Atwood number) will require further investigation of other strategies such as spatial shock shaping. Finally, more integrated scenarios involving sudden deceleration, such as in ICF, will have to employ a new class of optimization functions suitable for late-time RTI-induced mix, for example, by minimizing mass modulation.
\newline

\textit{Acknowledgements --} This work was supported by EPSRC and First Light Fusion under the AMPLIFI Prosperity Partnership - EP/X025373/1; the NNSA under DOE DE-NA0004148, Israel Science Foundation Grant No. 418/22, DOE Contract DE-AC52-07NA27344 and the LLNL-LDRD Program Project No. 21-SI-006. The research was also supported by the ESRF beamtime SC-5443. The Lawrence Livermore National Laboratory journal number is LLNL-JRNL-2010870.
\clearpage

\appendix

\section{Dynamic compression experiment}
The dynamic compression experiment was performed at the European Synchrotron using the ``Grengen" pulsed power generator charged to $+45~\SI{}{\kilo\volt}$ \cite{Maler2024}. In the experiment, the generator delivered a current pulse of $\sim110~\SI{}{\kilo\ampere}$ in $\sim600~\SI{}{\nano\second}$ into a copper strip encased in dielectric tamp, driving the resistive explosion. The electrical current discharge was measured by a custom-built self-integrating Rogowski coil, and the voltage was measured by a P6015A Tektronix high voltage divider. From the electrical signals, we estimate the total energy deposited into the into the metallic foil to be $\sim1~\SI{}{\kilo\joule}$ in approximately $\sim 250~\SI{}{\nano\second}$ as shown in Supplementary Figure A.  

The exploding copper strip was $10~\SI{}{\milli\meter}$ wide, $40~\SI{}{\milli\meter}$ long, and $12.5~\SI{}{\micro\meter}$ in thickness. The length and cross-section were chosen by matching the specific current action integral to obtain an almost critically damped discharge \cite{Oreshkin2020}. This discharge is characterized by the highest energy coupling efficiency and rate of energy density deposition. At peak current, the copper metal was expected to reach temperatures $>10~\SI{}{\kilo\kelvin}$ and consequently thermal pressures in the multi-$\SI{}{\giga\pascal}$ regime. The high thermal pressure of the exploded foil led to its expansion and a shock wave launch into the dielectric material that was surrounding it. Magnetohydrodynamic simulations of exploding wire arrays performed in the code GORGON estimated peak pressure of the fully formed shock wave at $\sim 500~\SI{}{\mega\pascal}$ followed by exponential decay to a background of $\sim 100~\SI{}{\mega\pascal}$ over $\sim5~\SI{}{\micro\second}$ \cite{10.1063/5.0272963}. More details related to this technique and specific driver are given in recent publications \cite{Strucka2023,Maler2024}.

\section{Multi-frame X-ray imaging}
The hydrodynamic evolution of the system was measured using ultra-high-speed synchrotron-based X-ray imaging. Approximate layout of the experiment is shown in Supplementary Figure B. The electron bunches stored in the storage ring of the synchrotron passed every $176~\SI{}{\nano\second}$ through two axially aligned long-period undulators ($\lambda_u = 32~\SI{}{\milli\meter}$). The undulators produced polychromatic X-ray radiation with spectrum spanning $20-50~\SI{}{\kilo\electronvolt}$ (characteristic energy $\tilde{E} \approx 30~\SI{}{\kilo\electronvolt}$). The X-rays propagated for $150~\SI{}{\meter}$ in a vacuum tube before passing through a spectral filter consisting of $1~\SI{}{\milli\meter}$ thick Beryllium and $\sim0.8~\SI{}{\milli\meter}$ thick Diamond window. The beam then passed through the shock-compressed sample, and another $9~\SI{}{\meter}$ before impinging on a LYSO:Ce X-ray scintillator (40 ns decay time, Hilger Crystals, UK) that converted the X-ray pulse into visible light. Due to the long propagation distances, the resulting radiographs contained both absorption-based features and edge-enhancement due to propagation-based phase-contrast imaging. 

The optical emission from the crystal scintillators was relayed via two pellicle beamsplitters to two Shimadzu HPV-X2 fast-framing cameras operating at $5~\SI{}{\mega\hertz}$. To capture every X-ray pulse, the cameras operated in parallel with a $100~\SI{}{\nano\second}$ offset. The resulting imaging could obtain up to $256$ consecutive radiographs with an effective interframe time of $176~\SI{}{\nano\second}$ ($5.68~\SI{}{\mega\hertz}$), exposure time of $100~\SI{}{\nano\second}$, $8\times12.8~\SI{}{\milli\meter\squared}$ field-of-view, and $32~\SI{}{\micro\meter}$ resolution.

Throughout the experiments, the X-ray beam was gated by a fast shutter that opened $\sim100~\si{\micro\second}$ prior to the start of measurement. For samples with high absorption contrast, this pre-illumination by multiple synchrotron pulses left a residual, spatially non-uniform conversion-efficiency imprint visible in early frames, as is visible in Fig.~\ref{fig_exp}(b). Additional details regarding the imaging setup can be found in previous publications \cite{Strucka2023,31,32}.

\section{Sample design and manufacture}
All instability samples were manufactured out of gelatine slabs with moulded sinusoidal perturbations (as shown in Figure~\ref{scheme}(a) in turquoise color). The mass density of the gelatine was $\sim 1.03~\SI{}{\gram\per\centi\meter\cubed}$ and yield strength was approximately $Y\sim 10~\SI{}{\kilo\pascal}$~\cite{Bakhrakh1997}. The benefit presented by using gelatine was the simplicity of manufacture of arbitrary shapes using prefabricated metallic / 3D printed plastic moulds, sufficiently high strength to support fixed geometries, but simultaneously sufficiently low strength such that the material behaves hydrodynamically in the experiment.

In all experiments, the sinusoidal interface was manufactured with a wavelength of $\lambda = 0.95\pm0.05~\SI{}{\milli\meter}$ and initial amplitude $a_0 = 0.1\pm0.02~\SI{}{\milli\meter}$, setting the dimensionless amplitude to $\left(2\pi / \lambda \right)a_0 \approx 0.66$; a regime where the initial dynamics can be weakly nonlinear. Additionally, the Atwood number of the interface between the gelatine ($\rho_0 = \rho_g = 1.03 ~ \SI{}{\gram\per\centi\meter\cubed}$) and the air ($\rho_1 = \rho_\text{air} = 0.0013~\SI{}{\gram\per\centi\meter\cubed}$) is $A=-1$, setting the initial conditions in a regime of the parameter space where the resulting dynamics should involve strongly asymmetric peak and trough growth and rapid jetting. 

In both the baseline and the suppression geometry, shown in Figure \ref{scheme}(b), the sinusoidal interface was positioned at a distance of $\sim5~\SI{}{\milli\meter}$ from the silicone layer and $\sim 6~\SI{}{\milli\meter}$ from the exploding copper foil. 

The only hydrodynamically relevant modification to the suppression geometry consisted of introducing computationally optimized voids behind the interface. The void geometry shown in Figure \ref{scheme}(b)(ii) was used in the experiment. These voids formed a periodic structure repeating every $1.9~\SI{}{\milli\meter}$, with each void described by the following linear functions:

\begin{equation}
I(x) = \begin{cases}
0.5 &\text{for $0\leq x<0.013 \SI{}{\centi\meter}$}\\
0.5 + 1.67 \left(x-0.013\right)&\text{for $0.013\leq x<0.031~\SI{}{\centi\meter}$}\\
0.53 + 0.79 \left(x-0.031\right)&\text{for $0.031\leq x<0.073~\SI{}{\centi\meter}$}\\
0.56 + 0.63 \left(x-0.073\right)&\text{for $0.073\leq x<0.081~\SI{}{\centi\meter}$}\\
0.57 + 0.43 \left(x-0.081\right)&\text{for $0.081\leq x<0.095~\SI{}{\centi\meter}$}.\\
\end{cases}
\end{equation}

This was an approximation of the optimal design which was prescribed by three piece-wise cubic Hermite interpolating polynomials that are fully defined by the following four points:
\begin{equation}
\begin{aligned}
    (x_0, y_0) &= (0.0,0.5)~\SI{}{\centi\meter},\\
    (x_1, y_1) &= (0.03167, 0.5014)~\SI{}{\centi\meter},\\ (x_2, y_2) &= (0.06333, 0.5588)~\SI{}{\centi\meter},\\(x_3, y_3) &= (0.095, 0.5716)~\SI{}{\centi\meter}.   
\end{aligned}
\end{equation}
The optimal void geometry was found by using the MARBL hydrodynamics code in combination with trust region Bayesian optimization algorithm. The objective function optimized by the algorithm was the maximum distance between the peaks and the troughs over the entirety of the simulation. This objective function assured that no significant spike growth occurred at any step of the simulation. In total, only $240$ runs were required to arrive at the optimal geometry.

\section{Hydrodynamic simulations}
All simulated results shown in this publication were obtained by two-dimensional simulations using the MARBL hydrodynamics code; an arbitrary Lagrangian-Eulerian code developed at Lawrence Livermore National Laboratory that uses high-order finite elements to numerically solve conservation equations \cite{Anderson2020,Rieben2020,Anderson2018}. High-order finite elements are useful for increasing accuracy of the simulations - for instance - by more accurately accounting for deformation at material interfaces. Each of our optimization simulation runs included approximately 16000 elements and was ran for $11~\SI{}{\micro\second}$, with the final design re-ran at a higher resolution with 60000 elements. A similar simulation method was used in previous work by Sterbentz \cite{Sterbentz2022} and compared against the scaling laws developed by Dimonte \cite{Dimonte2011}.

We used the Livermore Equation of State (LEOS) library to model the materials in our simulations. As a substitute for gelatine, we used the equation of state of water (dominant component of gelatine, table LEOS 2010). The main discrepancy between these two materials is the slightly higher density of gelatine ($\approx 1.03~\SI{}{\gram\per\centi\meter^3}$). At these low concentrations of gelatine ($3\%$), material strength effects are expected to be minor. Additionally, even if the gelatine is initially solid in the experiment, it is known that at pressures $\gg Y \approx 10~\SI{}{\kilo\pascal}$ the gelatine bonds break and the sample liquefies. As such, we typically neglected most strength effects in our simulations with an exception of a simple failure model for the gelatine. This model assumed that failure would occur if compressive stresses were equal to $-(2/3)Y_0$, where we selected $Y_0 = 100~\SI{}{\kilo\pascal}$ to match the void structure in the recorded radiographs. Importantly, the jet tip velocity and mean speed in the simulations were largely unaffected by this choice. Material properties of air were modelled using the LEOS 2260 equation of state.

\nocite{*}

\bibliography{apssamp}

@article{Sano2020,
  title = {Suppression of the Richtmyer-Meshkov instability due to a density transition layer at the interface},
  volume = {102},
  ISSN = {2470-0053},
  url = {http://dx.doi.org/10.1103/PhysRevE.102.013203},
  DOI = {10.1103/physreve.102.013203},
  number = {1},
  journal = {Physical Review E},
  publisher = {American Physical Society (APS)},
  author = {Sano,  Takayoshi and Ishigure,  Kazuki and Cobos-Campos,  Francisco},
  year = {2020},
  month = jul 
}

@article{Liang2021,
  title = {Shock-induced dual-layer evolution},
  volume = {929},
  ISSN = {1469-7645},
  url = {http://dx.doi.org/10.1017/jfm.2021.903},
  DOI = {10.1017/jfm.2021.903},
  journal = {Journal of Fluid Mechanics},
  publisher = {Cambridge University Press (CUP)},
  author = {Liang,  Yu and Luo,  Xisheng},
  year = {2021},
  month = nov 
}

@article{Vazirani2024,
  title = {Bayesian batch optimization for molybdenum versus tungsten inertial confinement fusion double shell target design},
  volume = {17},
  ISSN = {1932-1872},
  url = {http://dx.doi.org/10.1002/sam.11698},
  DOI = {10.1002/sam.11698},
  number = {3},
  journal = {Statistical Analysis and Data Mining: The ASA Data Science Journal},
  publisher = {Wiley},
  author = {Vazirani,  Nomita N. and Sacks,  Ryan and Haines,  Brian M. and Grosskopf,  Michael J. and Stark,  David J. and Bradley,  Paul A.},
  year = {2024},
  month = jun 
}

@article{10.1063/5.0272963,
    author = {Bokman, Guillaume T. and Fiorini, Samuele and Strucka, Jergus and Lukić, Bratislav and Bland, Simon N. and Mughal, Kassim and Liu, Siwei and Rack, Alexander and Supponen, Outi},
    title = {Planar shock-induced bubble collapse and jetting in water captured via x-ray phase contrast imaging},
    journal = {Applied Physics Letters},
    volume = {127},
    number = {1},
    pages = {014102},
    year = {2025},
    month = {07},
    issn = {0003-6951},
    doi = {10.1063/5.0272963},
    url = {https://doi.org/10.1063/5.0272963}
    }

@article{Schill2023,
  title = {Suppression of Richtmyer-Meshkov Instability via Special Pairs of Shocks and Phase Transitions},
  author = {Schill, W. J. and Armstrong, M. R. and Nguyen, J. H. and Sterbentz, D. M. and White, D. A. and Benedict, L. X. and Rieben, R. N. and Hoff, A. and Lorenzana, H. E. and Belof, J. L. and La Lone, B. M. and Staska, M. D.},
  journal = {Phys. Rev. Lett.},
  volume = {132},
  issue = {2},
  pages = {024001},
  numpages = {5},
  year = {2024},
  month = {Jan},
  publisher = {American Physical Society},
  doi = {10.1103/PhysRevLett.132.024001},
  url = {https://link.aps.org/doi/10.1103/PhysRevLett.132.024001}
}

@article{Samtaney2003,
  title = {Suppression of the Richtmyer–Meshkov instability in the presence of a magnetic field},
  volume = {15},
  ISSN = {1089-7666},
  url = {http://dx.doi.org/10.1063/1.1591188},
  DOI = {10.1063/1.1591188},
  number = {8},
  journal = {Physics of Fluids},
  publisher = {AIP Publishing},
  author = {Samtaney,  Ravi},
  year = {2003},
  month = jun,
  pages = {L53–L56}
}

@article{Sano2013,
  title = {Critical Magnetic Field Strength for Suppression of the Richtmyer-Meshkov Instability in Plasmas},
  volume = {111},
  ISSN = {1079-7114},
  url = {http://dx.doi.org/10.1103/PhysRevLett.111.205001},
  DOI = {10.1103/physrevlett.111.205001},
  number = {20},
  journal = {Physical Review Letters},
  publisher = {American Physical Society (APS)},
  author = {Sano,  Takayoshi and Inoue,  Tsuyoshi and Nishihara,  Katsunobu},
  year = {2013},
  month = nov 
}

@article{Sterbentz2022,
  title = {Design optimization for Richtmyer–Meshkov instability suppression at shock-compressed material interfaces},
  volume = {34},
  ISSN = {1089-7666},
  url = {http://dx.doi.org/10.1063/5.0100100},
  DOI = {10.1063/5.0100100},
  number = {8},
  journal = {Physics of Fluids},
  publisher = {AIP Publishing},
  author = {Sterbentz,  Dane M. and Jekel,  Charles F. and White,  Daniel A. and Aubry,  Sylvie and Lorenzana,  Hector E. and Belof,  Jonathan L.},
  year = {2022},
  month = aug 
}

@article{Zylstra2022,
  title = {Burning plasma achieved in inertial fusion},
  volume = {601},
  ISSN = {1476-4687},
  url = {http://dx.doi.org/10.1038/s41586-021-04281-w},
  DOI = {10.1038/s41586-021-04281-w},
  number = {7894},
  journal = {Nature},
  publisher = {Springer Science and Business Media LLC},
  author = {Zylstra,  A. B. and Hurricane,  O. A. and Callahan,  D. A. and Kritcher,  A. L. and Ralph,  J. E. and Robey,  H. F. and Ross,  J. S. and Young,  C. V. and Baker,  K. L. and Casey,  D. T. and D\"{o}ppner,  T. and Divol,  L. and Hohenberger,  M. and Le Pape,  S. and Pak,  A. and Patel,  P. K. and Tommasini,  R. and Ali,  S. J. and Amendt,  P. A. and Atherton,  L. J. and Bachmann,  B. and Bailey,  D. and Benedetti,  L. R. and Berzak Hopkins,  L. and Betti,  R. and Bhandarkar,  S. D. and Biener,  J. and Bionta,  R. M. and Birge,  N. W. and Bond,  E. J. and Bradley,  D. K. and Braun,  T. and Briggs,  T. M. and Bruhn,  M. W. and Celliers,  P. M. and Chang,  B. and Chapman,  T. and Chen,  H. and Choate,  C. and Christopherson,  A. R. and Clark,  D. S. and Crippen,  J. W. and Dewald,  E. L. and Dittrich,  T. R. and Edwards,  M. J. and Farmer,  W. A. and Field,  J. E. and Fittinghoff,  D. and Frenje,  J. and Gaffney,  J. and Gatu Johnson,  M. and Glenzer,  S. H. and Grim,  G. P. and Haan,  S. and Hahn,  K. D. and Hall,  G. N. and Hammel,  B. A. and Harte,  J. and Hartouni,  E. and Heebner,  J. E. and Hernandez,  V. J. and Herrmann,  H. and Herrmann,  M. C. and Hinkel,  D. E. and Ho,  D. D. and Holder,  J. P. and Hsing,  W. W. and Huang,  H. and Humbird,  K. D. and Izumi,  N. and Jarrott,  L. C. and Jeet,  J. and Jones,  O. and Kerbel,  G. D. and Kerr,  S. M. and Khan,  S. F. and Kilkenny,  J. and Kim,  Y. and Geppert Kleinrath,  H. and Geppert Kleinrath,  V. and Kong,  C. and Koning,  J. M. and Kroll,  J. J. and Kruse,  M. K. G. and Kustowski,  B. and Landen,  O. L. and Langer,  S. and Larson,  D. and Lemos,  N. C. and Lindl,  J. D. and Ma,  T. and MacDonald,  M. J. and MacGowan,  B. J. and Mackinnon,  A. J. and MacLaren,  S. A. and MacPhee,  A. G. and Marinak,  M. M. and Mariscal,  D. A. and Marley,  E. V. and Masse,  L. and Meaney,  K. and Meezan,  N. B. and Michel,  P. A. and Millot,  M. and Milovich,  J. L. and Moody,  J. D. and Moore,  A. S. and Morton,  J. W. and Murphy,  T. and Newman,  K. and Di Nicola,  J.-M. G. and Nikroo,  A. and Nora,  R. and Patel,  M. V. and Pelz,  L. J. and Peterson,  J. L. and Ping,  Y. and Pollock,  B. B. and Ratledge,  M. and Rice,  N. G. and Rinderknecht,  H. and Rosen,  M. and Rubery,  M. S. and Salmonson,  J. D. and Sater,  J. and Schiaffino,  S. and Schlossberg,  D. J. and Schneider,  M. B. and Schroeder,  C. R. and Scott,  H. A. and Sepke,  S. M. and Sequoia,  K. and Sherlock,  M. W. and Shin,  S. and Smalyuk,  V. A. and Spears,  B. K. and Springer,  P. T. and Stadermann,  M. and Stoupin,  S. and Strozzi,  D. J. and Suter,  L. J. and Thomas,  C. A. and Town,  R. P. J. and Tubman,  E. R. and Trosseille,  C. and Volegov,  P. L. and Weber,  C. R. and Widmann,  K. and Wild,  C. and Wilde,  C. H. and Van Wonterghem,  B. M. and Woods,  D. T. and Woodworth,  B. N. and Yamaguchi,  M. and Yang,  S. T. and Zimmerman,  G. B.},
  year = {2022},
  month = jan,
  pages = {542–548}
}

@article{Randolph2016,
  title = {Process Development and Micro-Machining of MARBLE Foam-Cored Rexolite Hemi-Shell Ablator Capsules},
  volume = {70},
  ISSN = {1943-7641},
  url = {http://dx.doi.org/10.13182/FST15-205},
  DOI = {10.13182/fst15-205},
  number = {2},
  journal = {Fusion Science and Technology},
  publisher = {Informa UK Limited},
  author = {Randolph,  Randall B. and Oertel,  John A. and Schmidt,  Derek W. and Lee,  Matthew N. and Patterson,  Brian M. and Henderson,  Kevin C. and Hamilton,  Christopher E.},
  year = {2016},
  month = sep,
  pages = {230–236}
}

@article{Kuczek2023,
  title = {Simulated impact of fill tube geometry on recent high-yield implosions at the National Ignition Facility},
  volume = {30},
  ISSN = {1089-7674},
  url = {http://dx.doi.org/10.1063/5.0156346},
  DOI = {10.1063/5.0156346},
  number = {9},
  journal = {Physics of Plasmas},
  publisher = {AIP Publishing},
  author = {Kuczek,  J. J. and Haines,  B. M.},
  year = {2023},
  month = sep 
}

@article{Strucka2023,
  title = {Synchrotron radiography of Richtmyer–Meshkov instability driven by exploding wire arrays},
  volume = {35},
  ISSN = {1089-7666},
  url = {http://dx.doi.org/10.1063/5.0144839},
  DOI = {10.1063/5.0144839},
  number = {4},
  journal = {Physics of Fluids},
  publisher = {AIP Publishing},
  author = {Strucka,  J. and Lukic,  B. and Koerner,  M. and Halliday,  J. W. D. and Yao,  Y. and Mughal,  K. and Maler,  D. and Efimov,  S. and Skidmore,  J. and Rack,  A. and Krasik,  Y. and Chittenden,  J. and Bland,  S. N.},
  year = {2023},
  month = apr 
}

@article{Weber2020,
  title = {Mixing in ICF implosions on the National Ignition Facility caused by the fill-tube},
  volume = {27},
  ISSN = {1089-7674},
  url = {http://dx.doi.org/10.1063/1.5125599},
  DOI = {10.1063/1.5125599},
  number = {3},
  journal = {Physics of Plasmas},
  publisher = {AIP Publishing},
  author = {Weber,  C. R. and Clark,  D. S. and Pak,  A. and Alfonso,  N. and Bachmann,  B. and Berzak Hopkins,  L. F. and Bunn,  T. and Crippen,  J. and Divol,  L. and Dittrich,  T. and Kritcher,  A. L. and Landen,  O. L. and Le Pape,  S. and MacPhee,  A. G. and Marley,  E. and Masse,  L. P. and Milovich,  J. L. and Nikroo,  A. and Patel,  P. K. and Pickworth,  L. A. and Rice,  N. and Smalyuk,  V. A. and Stadermann,  M.},
  year = {2020},
  month = mar 
}

@article{Baker2020,
  title = {Fill tube dynamics in inertial confinement fusion implosions with high density carbon ablators},
  volume = {27},
  ISSN = {1089-7674},
  url = {http://dx.doi.org/10.1063/5.0011385},
  DOI = {10.1063/5.0011385},
  number = {11},
  journal = {Physics of Plasmas},
  publisher = {AIP Publishing},
  author = {Baker,  K. L. and Thomas,  C. A. and Dittrich,  T. R. and Landen,  O. and Kyrala,  G. and Casey,  D. T. and Weber,  C. R. and Milovich,  J. and Woods,  D. T. and Schneider,  M. and Khan,  S. F. and Spears,  B. K. and Zylstra,  A. and Kong,  C. and Crippen,  J. and Alfonso,  N. and Yeamans,  C. B. and Moody,  J. D. and Moore,  A. S. and Meezan,  N. B. and Pak,  A. and Fittinghoff,  D. N. and Volegov,  P. L. and Hurricane,  O. and Callahan,  D. and Patel,  P. and Amendt,  P.},
  year = {2020},
  month = nov 
}

@article{Haines2019,
  title = {Computational study of instability and fill tube mitigation strategies for double shell implosions},
  volume = {26},
  ISSN = {1089-7674},
  url = {http://dx.doi.org/10.1063/1.5115031},
  DOI = {10.1063/1.5115031},
  number = {10},
  journal = {Physics of Plasmas},
  publisher = {AIP Publishing},
  author = {Haines,  Brian M. and Daughton,  W. S. and Loomis,  E. N. and Merritt,  E. C. and Montgomery,  D. S. and Sauppe,  J. P. and Kline,  J. L.},
  year = {2019},
  month = oct 
}

@article{Haan2011,
  title = {Point design targets,  specifications,  and requirements for the 2010 ignition campaign on the National Ignition Facility},
  volume = {18},
  ISSN = {1089-7674},
  url = {http://dx.doi.org/10.1063/1.3592169},
  DOI = {10.1063/1.3592169},
  number = {5},
  journal = {Physics of Plasmas},
  publisher = {AIP Publishing},
  author = {Haan,  S. W. and Lindl,  J. D. and Callahan,  D. A. and Clark,  D. S. and Salmonson,  J. D. and Hammel,  B. A. and Atherton,  L. J. and Cook,  R. C. and Edwards,  M. J. and Glenzer,  S. and Hamza,  A. V. and Hatchett,  S. P. and Herrmann,  M. C. and Hinkel,  D. E. and Ho,  D. D. and Huang,  H. and Jones,  O. S. and Kline,  J. and Kyrala,  G. and Landen,  O. L. and MacGowan,  B. J. and Marinak,  M. M. and Meyerhofer,  D. D. and Milovich,  J. L. and Moreno,  K. A. and Moses,  E. I. and Munro,  D. H. and Nikroo,  A. and Olson,  R. E. and Peterson,  K. and Pollaine,  S. M. and Ralph,  J. E. and Robey,  H. F. and Spears,  B. K. and Springer,  P. T. and Suter,  L. J. and Thomas,  C. A. and Town,  R. P. and Vesey,  R. and Weber,  S. V. and Wilkens,  H. L. and Wilson,  D. C},
  year = {2011},
  month = may 
}

@article{Anderson2018,
  title={High-order multi-material {ALE} hydrodynamics},
  author={Anderson, Robert W and Dobrev, Veselin A and Kolev, Tzanio V and Rieben, Robert N and Tomov, Vladimir Z},
  journal={SIAM J. Sci. Comput.},
  volume={40},
  number={1},
  pages={B32--B58},
  year={2018},
  publisher={SIAM}
}

@techreport{Anderson2020,
title={The Multiphysics on Advanced Platforms Project},
author={Anderson, R. and Black, A. and Blakeley, B. and Bleile, R. and Camier, J.-S. and Ciurej, J. and Cook, A. and Dobrev, V. and Elliott, N. and Grondalski, J. and Harrison, C. and Hornung, R. and Kolev, T. and Legendre, M. and Liu, W. and Nissen, W. and Olson, B. and Osawe, M. and Papadimitriou, G. and Pearce, O. and Pember, R. and Skinner, A. and Stevens, D. and Stitt, T. and Taylor, L. and Tomov, V. and Rieben, R. and Vargas, A. and Weiss, K. and White, D. and
Busby, L.},
year={2020},
number={Tech. rep., LLNL-TR-815869},
institution={Lawrence Livermore National Laboratory, Livermore, CA (United States)},
doi = {10.2172/1724326}
}

@inproceedings{Rieben2020,
author={Robert Rieben},
doi={10.13140/RG.2.2.12326.14403},
booktitle={Exascale Computing Project Annual Meeting},
title={Poster: {T}he {MARBL} multi-physics code},
year={2020},
month={},
volume={},
pages={} 
}

@article{Dimonte2011,
  title={Use of the {R}ichtmyer--{M}eshkov instability to infer yield stress at high-energy densities},
  author={Dimonte, Guy and Terrones, Guillermo and Cherne, Frank J and Germann, Timothy C and Dupont, Virginie and Kadau, Kai and Buttler, William T and Oro, David M and Morris, Chris and Preston, Dean L},
  journal={Phys. Rev. Lett.},
  volume={107},
  number={26},
  pages={264502},
  year={2011},
  publisher={APS}
}

@book{Bakhrakh1997,
  title = {Hydrodynamic instability in strong media},
  url = {http://dx.doi.org/10.2172/515973},
  DOI = {10.2172/515973},
  institution = {Office of Scientific and Technical Information (OSTI)},
  author = {Bakhrakh,  S.M. and Drennov,  O.B. and Kovalev,  N.P.},
  year = {1997},
  month = mar 
}

@article{31,
  doi = {10.1364/oe.25.013857},
  url = {https://doi.org/10.1364/oe.25.013857},
  year = {2017},
  month = jun,
  publisher = {The Optical Society},
  volume = {25},
  number = {12},
  pages = {13857},
  author = {Margie P. Olbinado and Xavier Just and Jean-Louis Gelet and Pierre Lhuissier and Mario Scheel and Patrik Vagovic and Tokushi Sato and Rita Graceffa and Joachim Schulz and Adrian Mancuso and John Morse and Alexander Rack},
  title = {{MHz} frame rate hard X-ray phase-contrast imaging using synchrotron radiation},
  journal = {Optics Express}
}

@article{32,
  doi = {10.1088/1361-6463/aaa2f2},
  url = {https://doi.org/10.1088/1361-6463/aaa2f2},
  year = {2018},
  month = jan,
  publisher = {{IOP} Publishing},
  volume = {51},
  number = {5},
  pages = {055601},
  author = {Margie P Olbinado and Valentina Cantelli and Olivier Mathon and Sakura Pascarelli and Joerg Grenzer and Alexander Pelka and Melanie Roedel and Irene Prencipe and Alejandro Laso Garcia and Uwe Helbig and Dominik Kraus and Ulrich Schramm and Tom Cowan and Mario Scheel and Pierre Pradel and Thibaut De Resseguier and Alexander Rack},
  title = {Ultra high-speed x-ray imaging of laser-driven shock compression using synchrotron light},
  journal = {Journal of Physics D: Applied Physics}
}

@article{Maler2024,
  title = {Multi frame radiography of supersonic water jets interacting with a foil target},
  volume = {135},
  ISSN = {1089-7550},
  url = {http://dx.doi.org/10.1063/5.0186659},
  DOI = {10.1063/5.0186659},
  number = {4},
  journal = {Journal of Applied Physics},
  publisher = {AIP Publishing},
  author = {Maler,  D. and Belozerov,  O. and Godinger,  A. and Efimov,  S. and Strucka,  J. and Yao,  Y. and Mughal,  K. and Lukic,  B. and Rack,  A. and Bland,  S. N. and Krasik,  Ya. E.},
  year = {2024},
  month = jan 
}

@article{Huntington2018,
  title = {Ablative stabilization of Rayleigh-Taylor instabilities resulting from a laser-driven radiative shock},
  volume = {25},
  ISSN = {1089-7674},
  url = {http://dx.doi.org/10.1063/1.5022179},
  DOI = {10.1063/1.5022179},
  number = {5},
  journal = {Physics of Plasmas},
  publisher = {AIP Publishing},
  author = {Huntington,  C. M. and Shimony,  A. and Trantham,  M. and Kuranz,  C. C. and Shvarts,  D. and Di Stefano,  C. A. and Doss,  F. W. and Drake,  R. P. and Flippo,  K. A. and Kalantar,  D. H. and Klein,  S. R. and Kline,  J. L. and MacLaren,  S. A. and Malamud,  G. and Miles,  A. R. and Prisbrey,  S. T. and Raman,  K. S. and Remington,  B. A. and Robey,  H. F. and Wan,  W. C. and Park,  H.-S.},
  year = {2018},
  month = may 
}

@article{Olson2020,
  title = {Development of the Marble experimental platform at the National Ignition Facility},
  volume = {27},
  ISSN = {1089-7674},
  url = {http://dx.doi.org/10.1063/5.0018819},
  DOI = {10.1063/5.0018819},
  number = {10},
  journal = {Physics of Plasmas},
  publisher = {AIP Publishing},
  author = {Olson,  R. E. and Murphy,  T. J. and Haines,  B. M. and Douglas,  M. R. and Albright,  B. J. and Gunderson,  M. A. and Kim,  Y. and Cardenas,  T. and Hamilton,  C. E. and Randolph,  R. B.},
  year = {2020},
  month = oct 
}

@article{Zhou2017c,
  title = {Rayleigh–Taylor and Richtmyer–Meshkov instability induced flow,  turbulence,  and mixing. II},
  volume = {723–725},
  ISSN = {0370-1573},
  url = {http://dx.doi.org/10.1016/j.physrep.2017.07.008},
  DOI = {10.1016/j.physrep.2017.07.008},
  journal = {Physics Reports},
  publisher = {Elsevier BV},
  author = {Zhou,  Ye},
  year = {2017},
  month = dec,
  pages = {1–160}
}

@article{Zhou2017,
  title = {Rayleigh–Taylor and Richtmyer–Meshkov instability induced flow,  turbulence,  and mixing. I},
  volume = {720–722},
  ISSN = {0370-1573},
  url = {http://dx.doi.org/10.1016/j.physrep.2017.07.005},
  DOI = {10.1016/j.physrep.2017.07.005},
  journal = {Physics Reports},
  publisher = {Elsevier BV},
  author = {Zhou,  Ye},
  year = {2017},
  month = dec,
  pages = {1–136}
}

@article{Sterbentz2023,
  title = {Linear shaped-charge jet optimization using machine learning methods},
  volume = {134},
  ISSN = {1089-7550},
  url = {http://dx.doi.org/10.1063/5.0156373},
  DOI = {10.1063/5.0156373},
  number = {4},
  journal = {Journal of Applied Physics},
  publisher = {AIP Publishing},
  author = {Sterbentz,  Dane M. and Jekel,  Charles F. and White,  Daniel A. and Rieben,  Robert N. and Belof,  Jonathan L.},
  year = {2023},
  month = jul 
}

@article{turbo,
  author  = {Eriksson, David and Pearce, Michael and Gardner, Jacob R. and Turner, Ryan D. and Poloczek, Matthias},
  title   = {Scalable Global Optimization via Local Bayesian Optimization},
  year    = {2019},
  journal = {arXiv e-prints},
  eprint  = {1910.01739},
  archivePrefix = {arXiv},
  primaryClass  = {cs.LG},
  doi     = {10.48550/arXiv.1910.01739},
  url     = {https://arxiv.org/abs/1910.01739}
}

@article{Oreshkin2020,
  title = {Wire Explosion in Vacuum},
  volume = {48},
  ISSN = {1939-9375},
  url = {http://dx.doi.org/10.1109/TPS.2020.2985100},
  DOI = {10.1109/tps.2020.2985100},
  number = {5},
  journal = {IEEE Transactions on Plasma Science},
  publisher = {Institute of Electrical and Electronics Engineers (IEEE)},
  author = {Oreshkin,  Vladimir I. and Baksht,  Rina B.},
  year = {2020},
  month = may,
  pages = {1214–1248}
}

@article{Keenan2020,
  title = {Shock-driven kinetic and diffusive mix in high-Z pusher ICF designs},
  volume = {27},
  ISSN = {1089-7674},
  url = {http://dx.doi.org/10.1063/1.5140361},
  DOI = {10.1063/1.5140361},
  number = {2},
  journal = {Physics of Plasmas},
  publisher = {AIP Publishing},
  author = {Keenan,  Brett D. and Taitano,  William T. and Simakov,  Andrei N. and Chacón,  Luis and Albright,  Brian J.},
  year = {2020},
  month = feb 
}

@article{Prime2024,
  title = {Multiscale Richtmyer-Meshkov instability experiments to isolate the strain rate dependence of strength},
  volume = {109},
  ISSN = {2470-0053},
  url = {http://dx.doi.org/10.1103/PhysRevE.109.015002},
  DOI = {10.1103/physreve.109.015002},
  number = {1},
  journal = {Physical Review E},
  publisher = {American Physical Society (APS)},
  author = {Prime,  Michael B. and Fensin,  Saryu J. and Jones,  David R. and Dyer,  Joshua W. and Martinez,  Daniel T.},
  year = {2024},
  month = jan 
}

@article{PhysRevA.31.410,
  title = {Richtmyer-Meshkov instabilities in stratified fluids},
  author = {Mikaelian, Karnig O.},
  journal = {Phys. Rev. A},
  volume = {31},
  issue = {1},
  pages = {410--419},
  numpages = {0},
  year = {1985},
  month = {Jan},
  publisher = {American Physical Society},
  doi = {10.1103/PhysRevA.31.410},
  url = {https://link.aps.org/doi/10.1103/PhysRevA.31.410}
}

@article{PhysRevLett.87.265001,
  title = {Direct Observation of Mass Oscillations Due to Ablative Richtmyer-Meshkov Instability in Plastic Targets},
  author = {Aglitskiy, Y. and Velikovich, A. L. and Karasik, M. and Serlin, V. and Pawley, C. J. and Schmitt, A. J. and Obenschain, S. P. and Mostovych, A. N. and Gardner, J. H. and Metzler, N.},
  journal = {Phys. Rev. Lett.},
  volume = {87},
  issue = {26},
  pages = {265001},
  numpages = {4},
  year = {2001},
  month = {Dec},
  publisher = {American Physical Society},
  doi = {10.1103/PhysRevLett.87.265001},
  url = {https://link.aps.org/doi/10.1103/PhysRevLett.87.265001}
}

@article{10.1063/1.1459459,
    author = {Aglitskiy, Y. and Velikovich, A. L. and Karasik, M. and Serlin, V. and Pawley, C. J. and Schmitt, A. J. and Obenschain, S. P. and Mostovych, A. N. and Gardner, J. H. and Metzler, N.},
    title = {Direct observation of mass oscillations due to ablative Richtmyer–Meshkov instability and feedout in planar plastic targets},
    journal = {Physics of Plasmas},
    volume = {9},
    number = {5},
    pages = {2264-2276},
    year = {2002},
    month = {05},
    issn = {1070-664X},
    doi = {10.1063/1.1459459},
    url = {https://doi.org/10.1063/1.1459459},

}

@article{10.1063/1.4764287,
    author = {Aglitskiy, Y. and Karasik, M. and Velikovich, A. L. and Serlin, V. and Weaver, J. L. and Kessler, T. J. and Nikitin, S. P. and Schmitt, A. J. and Obenschain, S. P. and Metzler, N. and Oh, J.},
    title = {Observed transition from Richtmyer-Meshkov jet formation through feedout oscillations to Rayleigh-Taylor instability in a laser target},
    journal = {Physics of Plasmas},
    volume = {19},
    number = {10},
    pages = {102707},
    year = {2012},
    month = {10},
    issn = {1070-664X},
    doi = {10.1063/1.4764287},
    url = {https://doi.org/10.1063/1.4764287},

}

@article{10.1063/1.4818280,
    author = {Igumenshchev, I. V. and Goncharov, V. N. and Shmayda, W. T. and Harding, D. R. and Sangster, T. C. and Meyerhofer, D. D.},
    title = {Effects of local defect growth in direct-drive cryogenic implosions on OMEGA},
    journal = {Physics of Plasmas},
    volume = {20},
    number = {8},
    pages = {082703},
    year = {2013},
    month = {08},
    issn = {1070-664X},
    doi = {10.1063/1.4818280},
    url = {https://doi.org/10.1063/1.4818280},
}

@article{PhysRevLett.125.055001,
  title = {Multimode Hydrodynamic Instability Growth of Preimposed Isolated Defects in Ablatively Driven Foils},
  author = {Zulick, C. and Aglitskiy, Y. and Karasik, M. and Schmitt, A. J. and Velikovich, A. L. and Obenschain, S. P.},
  journal = {Phys. Rev. Lett.},
  volume = {125},
  issue = {5},
  pages = {055001},
  numpages = {7},
  year = {2020},
  month = {Jul},
  publisher = {American Physical Society},
  doi = {10.1103/PhysRevLett.125.055001},
  url = {https://link.aps.org/doi/10.1103/PhysRevLett.125.055001}
}

@article{Christ2025,
  title = {Two-photon polymerization for inertial fusion energy target fabrication},
  volume = {131},
  ISSN = {1432-0630},
  url = {http://dx.doi.org/10.1007/s00339-025-08657-x},
  DOI = {10.1007/s00339-025-08657-x},
  number = {7},
  journal = {Applied Physics A},
  publisher = {Springer Science and Business Media LLC},
  author = {Christ,  Fabian and Schaumann,  Gabriel and Schott,  Nils and Vetter,  Johanna and Blaeser,  Andreas and Roth,  Markus},
  year = {2025},
  month = jun 
}

@article{10.1063/5.0107542,
    author = {Pandolfi, Silvia and Carver, Thomas and Hodge, Daniel and Leong, Andrew F. T. and Kurzer-Ogul, Kelin and Hart, Philip and Galtier, Eric and Khaghani, Dimitri and Cunningham, Eric and Nagler, Bob and Lee, Hae Ja and Bolme, Cindy and Ramos, Kyle and Li, Kenan and Liu, Yanwei and Sakdinawat, Anne and Marchesini, Stefano and Kozlowski, Pawel M. and Curry, Chandra B. and Decker, Franz-Joseph and Vetter, Sharon and Shang, Jessica and Aluie, Hussein and Dayton, Matthew and Montgomery, David S. and Sandberg, Richard L. and Gleason, Arianna E.},
    title = {Novel fabrication tools for dynamic compression targets with engineered voids using photolithography methods},
    journal = {Review of Scientific Instruments},
    volume = {93},
    number = {10},
    pages = {103502},
    year = {2022},
    month = {10},
    issn = {0034-6748},
    doi = {10.1063/5.0107542},
    url = {https://doi.org/10.1063/5.0107542},
}

@article{10.1063/5.0273572,
    author = {Parisuaña, C. and Valdivia, M. P. and Bouffetier, V. and Kurzer-Ogul, K. and Pérez-Callejo, G. and Bott-Suzuki, S. and Casner, A. and Christiansen, N. S. and Czapla, N. and Eder, D. and Galtier, E. and Glenzer, S. H. and Goudal, T. and Haines, B. M. and Hodge, D. and Ikeya, M. and Izquierdo, L. and Khaghani, D. and Kim, Y. and Klein, S. and Koniges, A. and Lee, H. J. and Leininger, M. and Leong, A. F. T. and Lester, R. S. and Makita, M. and Mancelli, D. and Martin, W. M. and Nagler, B. and Sandberg, R. L. and Truong, A. and Vescovi, M. and Gleason, A. E. and Kozlowski, P. M.},
    title = {Shock propagation in aerogel and TPP foams for inertial fusion energy target design},
    journal = {Physics of Plasmas},
    volume = {32},
    number = {8},
    pages = {082707},
    year = {2025},
    month = {08},
    issn = {1070-664X},
    doi = {10.1063/5.0273572},
    url = {https://doi.org/10.1063/5.0273572},
}

@article{10.1063/5.0272820,
    author = {Hodge, D. S. and Leong, A. F. T. and Kurzer-Ogul, K. and Pandolfi, S. and Montgomery, D. S. and Shang, J. and Aluie, H. and Marchesini, S. and Liu, Y. and Li, K. and Sakdinawat, A. and Galtier, E. C. and Nagler, B. and Lee, H. J. and Cunningham, E. F. and Carver, T. E. and Bolme, C. A. and Ramos, K. J. and Khaghani, D. and Kozlowski, P. M. and Gleason, A. E. and Sandberg, R. L.},
    title = {Single-shot in-line x-ray phase-contrast imaging of void-shockwave interactions in fusion energy materials},
    journal = {Physics of Plasmas},
    volume = {32},
    number = {8},
    pages = {083903},
    year = {2025},
    month = {08},
    issn = {1070-664X},
    doi = {10.1063/5.0272820},
    url = {https://doi.org/10.1063/5.0272820},
}

@article{Escauriza2020,
  title = {Collapse dynamics of spherical cavities in a solid under shock loading},
  volume = {10},
  ISSN = {2045-2322},
  url = {http://dx.doi.org/10.1038/s41598-020-64669-y},
  DOI = {10.1038/s41598-020-64669-y},
  number = {1},
  journal = {Scientific Reports},
  publisher = {Springer Science and Business Media LLC},
  author = {Escauriza,  E. M. and Duarte,  J. P. and Chapman,  D. J. and Rutherford,  M. E. and Farbaniec,  L. and Jonsson,  J. C. and Smith,  L. C. and Olbinado,  M. P. and Skidmore,  J. and Foster,  P. and Ringrose,  T. and Rack,  A. and Eakins,  D. E.},
  year = {2020},
  month = may 
}

@article{Ranjan2011,
  title = {Shock-Bubble Interactions},
  volume = {43},
  ISSN = {1545-4479},
  url = {http://dx.doi.org/10.1146/annurev-fluid-122109-160744},
  DOI = {10.1146/annurev-fluid-122109-160744},
  number = {1},
  journal = {Annual Review of Fluid Mechanics},
  publisher = {Annual Reviews},
  author = {Ranjan,  Devesh and Oakley,  Jason and Bonazza,  Riccardo},
  year = {2011},
  month = jan,
  pages = {117–140}
}

@article{Clark2024,
  title = {Modeling ablator defects as a source of mix in high-performance implosions at the National Ignition Facility},
  volume = {31},
  ISSN = {1089-7674},
  url = {http://dx.doi.org/10.1063/5.0200730},
  DOI = {10.1063/5.0200730},
  number = {6},
  journal = {Physics of Plasmas},
  publisher = {AIP Publishing},
  author = {Clark,  D. S. and Allen,  A. and Baxamusa,  S. H. and Biener,  J. and Biener,  M. M. and Braun,  T. and Davidovits,  S. and Divol,  L. and Farmer,  W. A. and Fehrenbach,  T. and Kong,  C. and Millot,  M. and Milovich,  J. and Nikroo,  A. and Nora,  R. C. and Pak,  A. E. and Rubery,  M. S. and Stadermann,  M. and Sterne,  P. and Weber,  C. R. and Wild,  C.},
  year = {2024},
  month = jun 
}

\end{document}